\documentclass[submission,copyright,creativecommons]{eptcs}
\providecommand{\event}{MARS 2020} 
\usepackage{mathtools}
\usepackage{standalone}
\usepackage{amsmath, amsfonts, amssymb}
\usepackage{graphicx}

\usepackage{underscore}           
\usepackage{wrapfig} 
\usepackage{xcolor}
\usepackage{graphicx}
\usepackage{tabularx}
\usepackage{multirow}
\usepackage{nicefrac}
\usepackage[para]{footmisc}
\usepackage{xspace}
\graphicspath{{images/}{../images/}}
\usepackage{multirow}
\newcommand{\figseq}{\textsc{FigSeq}\xspace}
\newcommand{\storm}{\textsc{Storm}\xspace} 
\newcommand{\yams}{\textsc{Yams}\xspace} 
\newcommand{\riskspec}{\textsc{RiskSpectrum}\xspace} 
\newcommand{\riskspecPSA}{\textsc{RiskSpectrum PSA}\xspace} 
\newcommand{\kb}{\textsc{KB3}\xspace}
\usepackage{blindtext}
\usepackage{subfiles} 

\title{Various Ways to Quantify BDMPs
}
\author{Marc Bouissou
\institute{EDF-R\&D\\
Electricit\'e de France\\
Palaiseau, France}
\email{marc.bouissou@edf.fr}
\and
Shahid Khan \qquad Joost-Pieter Katoen
\institute{Software Modeling and Verification \\ 
RWTH Aachen University\\
Aachen, Germany}
\email{\{shahid.khan, katoen\}@cs.rwth-aachen.de}
\and 
Pavel Krcal
\institute{RiskSpectrum Software\\
Lloyd's Register\\
Stockholm, Sweden}
\email{pavel.krcal@lr.org}
}
\def\titlerunning{Various Ways to Quantify BDMPs}
\def\authorrunning{M. Bouissou, S. Khan, J.P. Katoen  \& P. Krcal}
\begin{document}
\maketitle

\begin{abstract}
A Boolean logic driven Markov process (BDMP) is a dependability analysis model that defines a continuous-time Markov chain (CTMC).
This formalism has high expressive power, yet it remains readable because its graphical representation stays close to standard fault trees.
The size of a BDMP is roughly speaking proportional to the size of the system it models, whereas the size of the CTMC specified by this BDMP suffers from exponential growth.
Thus quantifying large BDMPs can be a challenging task.
The most general method to quantify them is Monte Carlo simulation, but this may be intractable for highly reliable systems. On the other hand, some subcategories of BDMPs can be processed with much more efficient methods.
For example, BDMPs without repairs can be translated into dynamic fault trees, a formalism accepted as an input of the \storm model checker, that performs numerical calculations on sparse matrices, or they can be processed with the tool \figseq that explores paths going to a failure state and calculates their probabilities.
BDMPs with repairs can be quantified by \figseq (BDMPs capturing quickly and completely repairable behaviors are solved by a different algorithm), and by the I\&AB (Initiator and All Barriers) method, recently published and implemented in a prototype version of \riskspecPSA. This tool, based exclusively on Boolean representations looks for and quantifies minimal cut sets of the system, i.e., minimal combinations of component failures that induce the loss of the system.
This allows a quick quantification of large models with repairable components, standby redundancies and some other types of dependencies between components.
All these quantification methods have been tried on a benchmark whose definition was published at the MARS 2017 workshop: the model of emergency power supplies of a nuclear power plant.
In this paper, after a recall of the theoretical principles of the various quantification methods, we compare their performances on that benchmark.
\end{abstract}

\section{Introduction}
\label{sec:intro}

EDF has developed several methods and tools for creating and quantifying discrete stochastic models. 
In particular, Boolean logic driven Markov processes (BDMPs) are a powerful modeling formalism for the dependability analysis of dynamic systems~\cite{bouissou2003new}.
BDMPs have a graphical representation close to fault trees, widely used in safety and dependability studies of industrial systems. While standard fault trees are static models, in which basic events corresponding to  failure modes of components are supposed to be independent, in a BDMP, a single new graphical element, called trigger, can be used to specify various kinds of dependencies between basic events. In fact, a BDMP specifies a dynamic model: a (potentially very large) CTMC. A CTMC is a stochastic process $X_t$ where the random variable $X$ belongs to a discrete state space $D$ and $t$ is a continuous time scale. We will consider only the case of homogeneous Markov processes, characterized by the fact that $Pr(X_{t+s}= x_2 \mid X_t= x_1) = Pr(X_s= x_2 \mid X_0= x_1)$, where $X_0$ denotes the initial state of the process, at $t=0$, and $x_1$, $x_2$ are any states in $D$. 

\begin{wrapfigure}[16]{r}{0.3\textwidth}
\vspace{-0.5cm}
\begin{center}
{\includegraphics[clip, trim=0.05cm 0.02cm 0.01cm 0.01cm
width=0.3\textwidth]{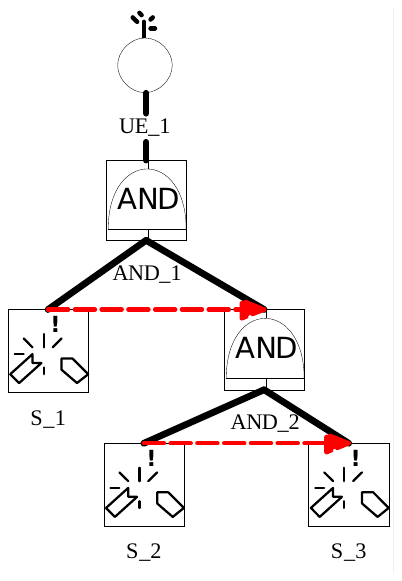}}
\end{center}
\caption{BDMP model.}
\label{fig:bdmp}
\end{wrapfigure}

For example, the system described at the beginning of Section~\ref{sec:clasicalmethods} (CTMC provided in Figure~\ref{fig:mc}) can be represented exactly by the BDMP of Figure~\ref{fig:bdmp}, containing two triggers: the upper trigger specifies that the subtree corresponding to the two backup components is not ``activated" as long as the first component is working, and thus cannot fail. The lower trigger specifies that, within this subtree, the component 3 is not activated as long as component 2 is working.

It is fairly easy to build BDMPs, but their quantification can require large resources, especially when they model highly reliable repairable systems. 
In this paper, we are going to examine various quantification methods and compare them on a benchmark case of medium complexity, a BDMP with 81 leaves representing the emergency power supplies of a nuclear power plant.

First, we will look at classical methods for the quantification of CTMCs. Given the context of the MARS workshop, the purpose is not to do a review of all existing methods potentially applicable to BDMPs, but rather to give a shortlist of methods supported by existing tools, so that the inputs and outputs of these tools can be put on the MARS model repository.

In addition to classical methods applicable to any CTMC, hence to BDMPs, we will present a method which can handle repairs in event tree/fault tree models and hence starting from a BDMP requires only a minimal translation effort: the I\&AB (Initiator and All Barriers) \cite{bouissouHernu2016} method.

This method provides a good compromise between the classical Probabilistic Safety Assessment (PSA) method used for safety analyses of nuclear power plants, which is exclusively based on Boolean models assuming the independence of basic events (except in very limited cases), and purely dynamic, state-space based methods, particularly those using analytical calculations on Markov chains. The price to pay is that only the temporal dependencies between initiators and failures of all remaining barriers are taken into account. Especially, restrictions on ordering of barrier failures which might be specified in a BDMP model are disregarded. The model analyzed by the I\&AB algorithm includes scenarios where all orderings of barrier failures are possible; they are quantified and counted in to the resulting failure frequency.
The principle of I\&AB consists in ``summarizing" the sequences leading to failure states of a Markov chain by cut sets, each of them containing one initiating event and failures of other components, or ``barriers."
Then a failure rate (usually called frequency in the PSA context) is calculated for each cut set, and the sum of these rates yields an equivalent failure rate for the whole system.
What makes this method so efficient is the fact that it uses closed-form formulae for cut sets quantification~\cite{bouissouHernu2016}. There is a prototype implementation of this method in the event/fault tree solver of \riskspec~\cite{RiskSpectrumTheoryManual}, a commercial software suite for safety analyses. Therefore, the I\&AB analysis can benefit from a state-of-the-art solver, efficiently decomposing probabilistic Boolean models into minimal cut sets. We show on the benchmark case that, although less precise than other methods, I\&AB can give a fair lower bound for the system reliability and a good insight into dominating failure scenarios: these two characteristics are needed in the evaluation of safety critical systems.

The paper is organized as follows: Sections \ref{sec:clasicalmethods} and \ref{sec:InABmethd} explain the principles of the different quantification methods. Then the tools for implementing these methods are presented in Section \ref{sec:tools}. The rest of the paper is dedicated to the case study beginning with a short recall of the benchmark definition in Section \ref{sec:casestudy}. The performance comparison of various quantification methods in the context of repairable and non-repairable version is presented in Section \ref{subsec:comparison:Repariable} and \ref{subsec:comparison:Non-Repariable}, respectively. Section \ref{sec:conc} concludes the work with some future directions.

\section{Classical Methods for Markov Chains}
\label{sec:clasicalmethods}

In order to illustrate differences between the calculation methods quoted in this section and the next one dedicated to I\&AB, we will take a very simple example (BDMP provided in Figure~\ref{fig:bdmp}) and show the interpretation of the methods in terms of state graphs. Let us consider a system consisting of three components $S1$, $S2$, and $S3$, with constant failure and repair rates ($\lambda_i,\mu_i$). In the perfect state, component $S1$ is in operation and components $S2$ and $S3$ are in standby. As soon as $S1$ fails, $S2$ starts functioning. When both $S1$ and $S2$ are failed, $S3$ replaces them. Note that this little system could also represent a cut set of a large system. 

\begin{wrapfigure}[11]{r}{0.4\textwidth}
\vspace{-0.8cm}
\centering
{\includegraphics[clip, trim=0.7cm 9.8cm 10.5cm 0.9cm,width=0.4\textwidth]{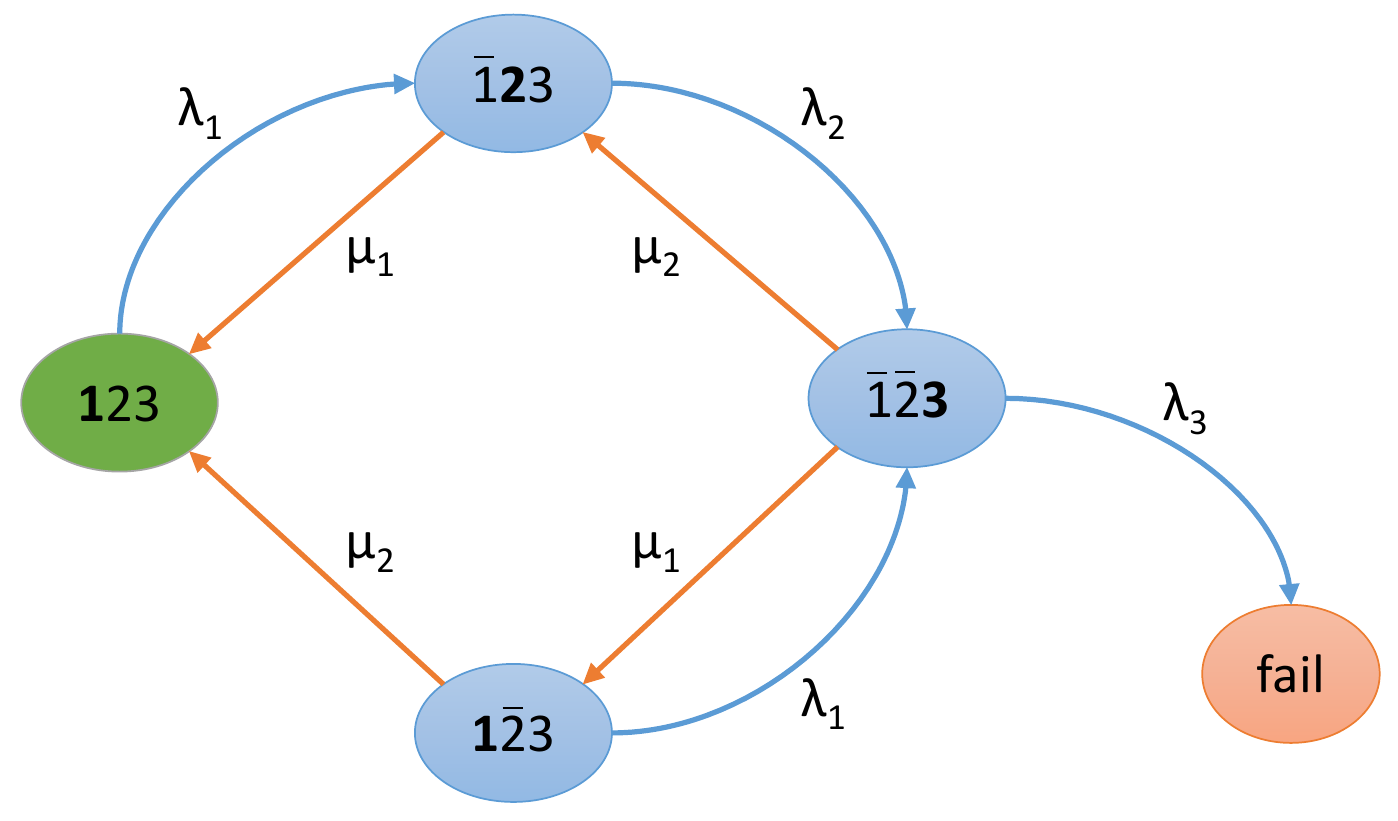}}
\caption{CTMC modelling the system.}
\label{fig:mc}
\end{wrapfigure}

A truly dynamic model, that takes all these hypotheses into account, is the CTMC of Figure~\ref{fig:mc}. We use a bold font for components in operation and a regular font for those in standby; numbers with bars indicate failed components.
Notice that there is only one functioning component in each state. 
In this Markov chain, the only possible path (without loop) resulting in the top event $UE\_1$ is successive failures of S1, S2, and S3. Notice we do not explore further after the first failure, therefore no transition going out of \textit{fail} state is provided.

\subsection{From a Matrix Representation}
The \storm model checker \cite{DBLP:journals/corr/abs-2002-07080} uses \emph{numerical} methods to compute metrics of CTMCs either using a sparse matrix or symbolic representation.
The symbolic methods are based on binary decision diagrams (BDDs). BDD-based analysis of dynamic fault trees (DFTs) using \storm is a topic of further research. As we use the sparse engine of \storm for DFT analysis, we briefly outline the matrix-based solution of Figure~\ref{fig:mc}.
In a CTMC, we have a rate function $r(s)=\sum_{s^\prime \in S}{\textbf{R}(s,s^{\prime})}$ (where $S$ is the set of CTMC states and $\textbf{R}(s,s^{\prime})$ is the transition rate of moving from state $s$ to $s^\prime$) associated to each state $s$ (see \cite{DBLP:series/natosec/Katoen13} for details).
This means the probability to wait $t$ time units in state $s$ is $1-e^{-r(s)\cdot{t}}$. Thus the average state residence time is $\frac{1}{r(s)}$. The rates for each state of Figure~\ref{fig:mc} are $r(\textbf{1}23)=\lambda_1$, $r(\bar{1}\textbf{2}3)=\mu_1+\lambda_2$,
$r(\bar{1}\bar{2}\textbf{3})=\mu_2+\mu_1+\lambda_3$,
$r(\textbf{1}\bar{2}3)=\mu_2+\lambda_1$, and 
$r(fail)=0$. Given $r(s)$ and transition rate $\textbf{R}(s,s^{\prime})$, we define the transition probability function $\textbf{P}(s,s^\prime)$ as: $\textbf{P}(s,s^{\prime})=\frac{\textbf{R}(s,s^\prime)}{r(s)}$. The $\textbf{R}(s,s^\prime)$ of Figure~\ref{fig:mc} is: \\[1ex]
\resizebox{0.98\hsize}{!}{
\begin{math}
\underbrace{
\begin{pmatrix}
0 &  \lambda_1  & 0 & 0 & 0\\
 \mu_1  & 0 &  \lambda_2  & 0& 0\\
0 &  \mu_2  & 0 &  \mu_1  &  \lambda_3 \\
 \mu_2  & 0 & 0 &  \lambda_1  & 0\\
0 & 0 & 0 & 0 & 0\\
\end{pmatrix}}_{R(s,s^\prime)}
 =
\underbrace{
\begin{pmatrix}
0 & 1 & 0 & 0 & 0\\
\nicefrac{\mu_1}{\mu_1 +\lambda_2} & 0 & \nicefrac{\lambda_2}{\mu_1 +\lambda_2} & 0& 0\\
0 & \nicefrac{\mu_2}{\mu_2+\mu_1+ \lambda_3} & 0 & \nicefrac{\mu_1}{\mu_2+\mu_1+ \lambda_3} & \nicefrac{\lambda_3}{\mu_2+\mu_1+ \lambda_3}\\
\nicefrac{\mu_2}{\mu_2+ \lambda_1} & 0 & 0 & \nicefrac{\lambda_1}{\mu_2+ \lambda_1} & 0\\
0 & 0 & 0 & 0 & 1\\
\end{pmatrix}}_{P(s,s^\prime)}
\cdot
\underbrace{
\begin{pmatrix}
\lambda_1 \\
 \mu_1  + \lambda_2 \\
 \mu_2 + \mu_1 +  \lambda_3 \\
 \mu_2 +  \lambda_1  \\
0
\end{pmatrix}}_{r(s)}
\end{math}
}

Intuitively speaking, the CTMC starts in its initial state, i.e., $\textbf{1}23$. Upon entering any state, the state residence time is determined by an exponential distribution with rate $r(s)$. Upon leaving the state $s$, the probability to move to state $s^\prime$ is then given by $\textbf{P}(s,s^\prime)$.

The \emph{unreliability} metric on a CTMC $\mathit{C}$ can be viewed as time-constrained reachability $\Diamond^{\leq t} \mathit{fail}$. That is we are interested in the probability of reaching a state labelled \textit{fail} (see Figure~\ref{fig:mc}) within a deadline $t$. In order to compute this probability we make goal states absorbing and obtain CTMC $\mathit{C}[G]$ (in this notation, $\mathit{G}$ represents the set of goal states). The probability of reaching a goal state (\emph{fail} in our case) from state $s$ is computed as:\\[1ex]
\resizebox{0.98\hsize}{!}{
\begin{math}
\underbrace{P(s\models \Diamond^{\leq t} G)}_{\text{timed reachability in }\textit{C}} = \underbrace{P(s\models \Diamond^{= t} G)}_{\text{timed reachability in }\textit{C}[G]} = \sum_{s^\prime\in G}{p_{s^\prime}(t)} \text{ where $p_s(t)$ is probability of being in state $s$ at time $t$}
\end{math}}

This equation implies that the problem of timed-reachability is equivalent to computing the transient probability distribution in the adapted CTMC $\mathit{C}[G]$, i.e., the CTMC in which we lump the goal states into an absorbing state. In order to compute (in a numerically stable manner \cite{DBLP:series/natosec/Katoen13}) the transient probability on a CTMC we normalize it by making rates of each state equal to $r$, where $r = \max_{s}r(s)$. All the states $s^\prime$ having $r(s^\prime) < r$ are augmented with a self loop of transition rate ($r-r(s^\prime)$). If a self loop already exists, then its probability becomes $\frac{r(s^\prime)}{r}\cdot\textbf{P}(s^\prime,s^\prime) + (1-\frac{r(s^\prime)}{r})$. For no self-looping transitions $\bar{\textbf{{P}}}(s^\prime,s^{\prime\prime})=\frac{r(s^\prime)}{r}\cdot {\textbf{P}}(s^\prime,s^{\prime\prime})$. Once we get the normalized CTMC, we solve a system of linear differential equation to obtain the transient probabilities $\underline{p}(t) = \underline{p}(0)\cdot e^{-rt}\cdot\sum_{i=0}^{\infty}{\frac{{(r\cdot t)}^{i}}{i!}}\cdot \bar{\textbf{P}}^i$. Since $\bar{\textbf{P}}$ is a stochastic matrix, computing its $i^{th}$ exponent is numerically stable, see \cite{DBLP:series/natosec/Katoen13} for details. \storm solves the system of differential equations using {off-the-shelf} solvers (EIGEN, GMM++, Gaussian elimination and a native solver) and \emph{Fox-Glynn} computation of Poisson probabilities \cite{fox1988computing}. To summarize, \storm internally populates matrices as sparse matrix data structure and uses matrix-vector multiplications to yield the numeric value of interest. Notice that the transient probability vector $\underline{p}(t)$ contains probability associated to each state and we are interested in only one element of this vector corresponding to the \textit{fail}-labelled state. The result that \storm computes is exact up to given user-defined tolerance.

\subsection{Sequence Based Calculation}
The tool \figseq (cf. § 4.2 and 5.3) implements a search and quantification of sequences leading to a target state.
Two quite different algorithms are needed: one for repairable, and one for non-repairable models \cite{bonbouissou1992}. In \figseq these algorithms are respectively called \emph{NRI} (for no return to the initial state) and \emph{NS} (for normal sequences). In both cases, the memory consumption is modest, and can be controlled without hindering numerical results: memory is only needed to store the dominant failure sequences that will be sorted by decreasing probability and displayed in the output. 

\emph{NRI} can only give an upper bound of the unreliability \cite{boncollet1994}, but this upper bound is very close to the true value for quickly and completely repairable systems \cite{keilson1986}. The graph of Figure~\ref{fig:mc} represents a \textit{completely} repairable system (from any state other than the system failure state, it is possible to go back to the perfect state). It is also \textit{quickly} repairable if all repair rates are larger than failure rates by at least one order of magnitude. The upper bound given by \emph{NRI} can be written as $1-e^{-\Lambda \cdot \epsilon \cdot t}$, where $\Lambda$ is the sum of all rates of transitions leaving the initial state, and $\epsilon$ is the probability to go from the initial state to a failure state of the system without returning to the initial state (hence the name \emph{NRI}). $\epsilon$ is calculated from transition probabilities in the discrete embedded Markov chain of the Markov process. This amounts to neglecting completely the time spent in all degraded states. Note that the output file of \figseq associated to this paper also gives an approximation of the asymptotic unavailability of the system, thanks to an extension of the NRI algorithm described in \cite{bouissoulefebvre2002}.
\figseq is based on an algorithm that explores loops in the Markov graph only once, while taking into account in the calculation of $\epsilon$ the fact that these loops can be traversed any number of times. In the example of Figure~\ref{fig:bdmp}, \figseq would only explore the three sequences: $\lambda_1, \lambda_2, \lambda_3 ; \lambda_1, \lambda_2, \mu_2 ; \lambda_1, \lambda_2, \mu_1, \lambda_1$. Only the first one will appear in \figseq's results.

\emph{NS} is a ``brute force" exploration of sequences \cite{colletrenault1997}, where the tree of all sequences starting from the initial state and going to a failure state is exhaustively explored, whether they include loops or not. Moreover, the time spent in all states of a sequence is taken into account in the calculation of the probability of this sequence. For a large state graph, or even for a small one if it includes loops (like in Figure~\ref{fig:mc}), this exploration must be limited by \emph{cutoff criteria} (maximum length, minimum probability, maximum number of failures and/or repairs). The \emph{NS} algorithm gives an upper and a \textit{lower} bound for the unreliability. The upper bound is obtained by considering that incomplete sequences lead to a failure state. For the model of Figure~\ref{fig:mc}, \emph{NS} could for example yield the following framing for unreliability $Q(t)$: $Pr(\lambda_1, \lambda_2, \lambda_3)(t) < Q(t) < Pr(\lambda_1, \lambda_2, \lambda_3)(t)+ Pr(\lambda_1, \lambda_2, \mu_2)(t) + Pr(\lambda_1, \lambda_2, \mu_1)(t)$. A tight upper bound can be obtained only for a non-repairable system, or, if repairable for a small enough mission time, otherwise the number of sequences that must be explored and quantified explodes.  

\subsection{Monte Carlo simulation}
The tool \yams (cf. § 4.2 and 5.3) uses a classical ``event driven" Monte Carlo simulation \cite{bouissou2013MCacceleration}.
It simulates a (usually large) number of sample stories (also called in other contexts traces or trials) drawn at random and then makes statistics on these stories.
For example, the unreliability at time $t$ is the proportion of simulated stories in which a system failure happened before $t$. 
Monte Carlo methods may suffer from excessive simulation times due to the high reliability of the systems under study: a very large number of simulations is required to produce a meaningful statistical picture of the failure modes.
In such cases, the most-likely failure modes dominate, and it is difficult to see the contribution of the less-likely modes. This is illustrated in the benchmark results of Section \ref{subsec:comparison:Repariable}: among 20 million simulations, \yams could find only 70 sequences repeated at least twice. This is easily explained by the results of \figseq: the sequence of rank 11 (in the list sorted by decreasing probabilities) is already more than 10 times less probable than the first one. Sometimes the failure probability to quantify is so small, that the simple observation of this probability is already very costly. 

\yams is a general purpose tool applicable to very complex, often non Markovian models. This is why it does not include any variance reduction method that could speed up the calculations. Since BDMPs are Markovian and well structured models, it would certainly be possible to find good acceleration techniques for them.

\section{The I\&AB Method}
\label{sec:InABmethd}
Classical methods are limited, at least for repairable systems: the largest BDMPs ever processed with \figseq had about 300 basic events, and Monte Carlo simulation often leads to unacceptable CPU times for highly reliable systems. 

\begin{wrapfigure}[10]{r}{0.48\textwidth}
\vspace{-0.8cm}
\centering
{\includegraphics[clip, trim=0.8cm 9.6cm 8.3cm 0.7cm, width=0.48\textwidth]{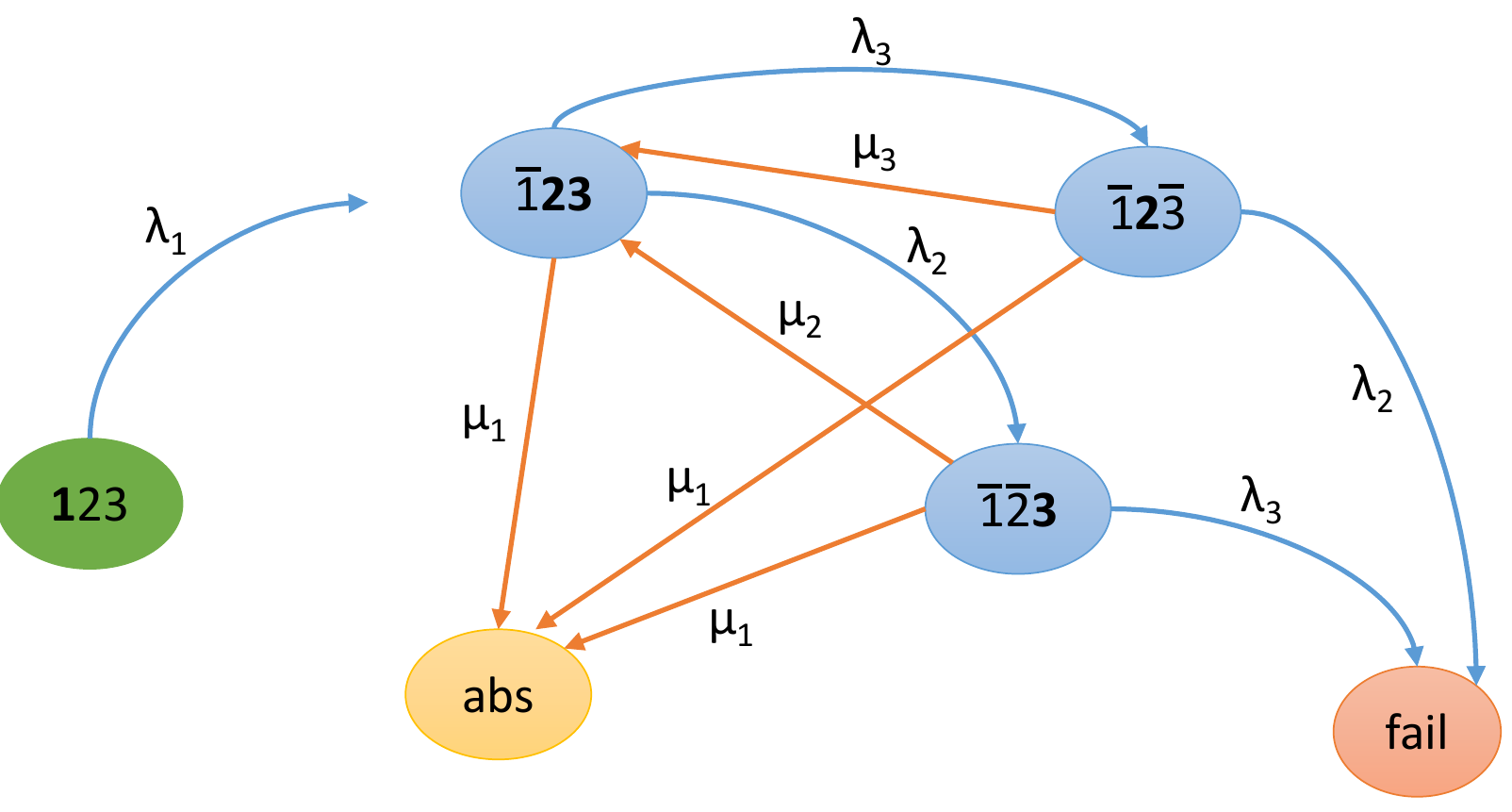}}
\caption{CTMC corresponding to I\&AB method.}
\label{fig:mc-exp}
\end{wrapfigure}

But, by using the ability of \kb \cite{bouissou2005yamsfigseq} to convert various models into static fault trees, it is possible to transform automatically a BDMP into a \riskspecPSA model suitable for quantification by I\&AB, which offers a novel way to quantify very large BDMPs.

The method was inspired by an insight of \cite{collet1995extension} noting that the majority of the behavior of a fully dynamic model is captured by the first-order relationships between the failure of functioning components (i.e. the frequency events) and the standby components which act as barriers to the initial failure (i.e. the remaining basic events in a cut set).

\noindent  The initiating event is modelled as a repairable event which fails with rate $\lambda_i$. 
While it is failed and under repair (with repair rate $\mu_i$), the barriers $B_1, B_2,\cdots, B_n$ are assumed to immediately begin to function. Barrier events may be either failure on-demand or failure in-function type events. 
Failure on-demand events may fail with probability $\gamma_j$ at the occurrence of the initiating event. 
If they fail, they are under repair (with repair rate $\mu_j$), and once repaired they cannot fail again (the system moves to a safe state). Failure in function events fail with rate $\lambda_k$ and are repaired with repair rate $\mu_k$. 
Failure in function events continue to operate after repair and may undergo successive cycles of failure and repair.
If at any stage the initiating event is repaired, the system moves to a ‘safe state’ and cannot fail until the next initiating event. 
These assumptions correspond to the state graph of Figure~\ref{fig:mc-exp} for the three-component system of Figure~\ref{fig:bdmp}. In this figure, there is a gap after the initial state's outgoing edge to show that the quantification by I\&AB is just the multiplication of the frequency of the initiating event by an (approximate) quantification of the probability that, starting from state $\bar{1}\textbf{23}$ the failure state \emph{fail} is reached, rather than the safe state \emph{abs}.   

For a large system, the I\&AB method requires the determination of cut sets in which the initiating event is distinguished from other events; each cut set is then quantified using the I\&AB method like for the little system described above. 
This method yields an analytic conservative approximation of the CTMC of Figure~\ref{fig:mc} for the cut set. 

The reader is referred to \cite{bouissouHernu2016, bouissou2018extensions} for a detailed description of the calculation. 
The paper \cite{bouissouHernu2016} gives the generic equations for the quantification of a cut set, and the procedure to obtain automatically the relevant cut sets from a BDMP, using the fault tree generation function of \kb. 
The paper \cite{bouissou2018extensions} gives the detailed analytical formulae for the quantification of a single cut set, and their extension in the case where there are deterministic delays.

\section{Tools Used for Our Experiments}
\label{sec:tools}
\subsection{Storm Model Checker}
\label{subsec:storm}
The process of probabilistic model checking (PMC)~\cite{DBLP:conf/lics/Katoen16} amounts to verifying a logic-based property against a state-space based description of the system. Therefore, the steps to enable PMC involve transforming the model of the system under study to an equivalent (up to an acceptable level of abstraction) state-space based description and writing properties (defined in an appropriate logic) for this system. The selection of the state-space model depends on the type of system (whether discrete, continuous or having non-determinism). Prevalent in the domain of PMC are discrete time Markov chains (DTMCs), continuous time Markov chains (CTMCs), Markov decision processes (MDPs), continuous time Markov decision processes (CTMDPs) and Markov automata (MA)~\cite{DBLP:conf/lics/Katoen16}, where Markov automaton (MA) is a very expressive probabilistic model subsuming DTMCs, CTMCs, MDPs, and CTMDPs. 

\storm~\footnote{\href{https://www.stormchecker.org/}{https://www.stormchecker.org/}}~\cite{DBLP:journals/corr/abs-2002-07080} is a state-of-the-art probabilistic model checker capable of model checking MA and its constituents. \storm is an open-source and freely available tool. It is quite competent and outperforms contemporary model checkers (on most of the benchmarks) as reported in QComp 2019~\footnote{\href{http://qcomp.org/competition/2019/}{http://qcomp.org/competition/2019/}}~\cite{Qcomp2019}. It uses numeric and symbolic methods for state-space generation. The architecture of \storm is modular and \textsc{StormPy}~\footnote{\href{https://moves-rwth.github.io/stormpy/}{https://moves-rwth.github.io/stormpy/}} (Python bindings of \storm) can be used for reconfiguration of \storm underlying engines/solvers and thus a quick prototyping of solution according to user requirements is possible. \storm can parse models specified in various modeling languages, e.g. dynamic fault trees (DFTs), GSPNs, PRISM, pGCL, JANI and DOT (explicit) format.

\storm supports DFT analysis and accepts DFT models in Galileo~\cite{sullivan1999galileo} or JSON format. Internally, \storm converts the DFT into an MA. The non-determinism is used to consider \textsf{SPARE} races. Once the underlying state-space of the DFT is available, all kinds of measures expressible in probabilistic temporal logic can be computed. This includes reliability, MTTF, conditional failure probabilities, etc. Availability is not covered as this metric involves repairs and the current implementation of \storm does not support repairable DFTs.

The first attempt to enable BDMPs analysis with \storm is through DFTs \cite{khanprdc}. In this approach, nine transformation rules were identified to convert a given BDMP into a DFT and then \storm DFT support was used to compute reliability metrics of the DFT. The main limitation of this work is that it focuses on non-repairable BDMPs. A recent work considers the conversion of repairable BDMP into \storm inputs, with two compositional approaches. The first one is based on Markov Automata, and the second one on generalized stochastic Petri nets (GSPN). \storm supports the JANI format~\cite{DBLP:conf/tacas/BuddeDHHJT17} which is a JSON based model interchange language to enable portability among different formal analysis tools. A BDMP with repairs can be transformed into Markov Automata that are written in JANI format. It can also be translated into a GSPN, another formalism natively supported by \storm. The transformation tools are of prototypical nature. Future work will focus on efficient state-space generation for repairable BDMPs.

\subsection{The KB3 Workbench}
\label{subsec:kb3}
\begin{wrapfigure}[18]{r}{0.6\textwidth}
\vspace{-0.3cm}
\centering
{\includegraphics[clip, trim=0.6cm 0.5cm 0.0cm 0.5cm, width=0.6\textwidth]{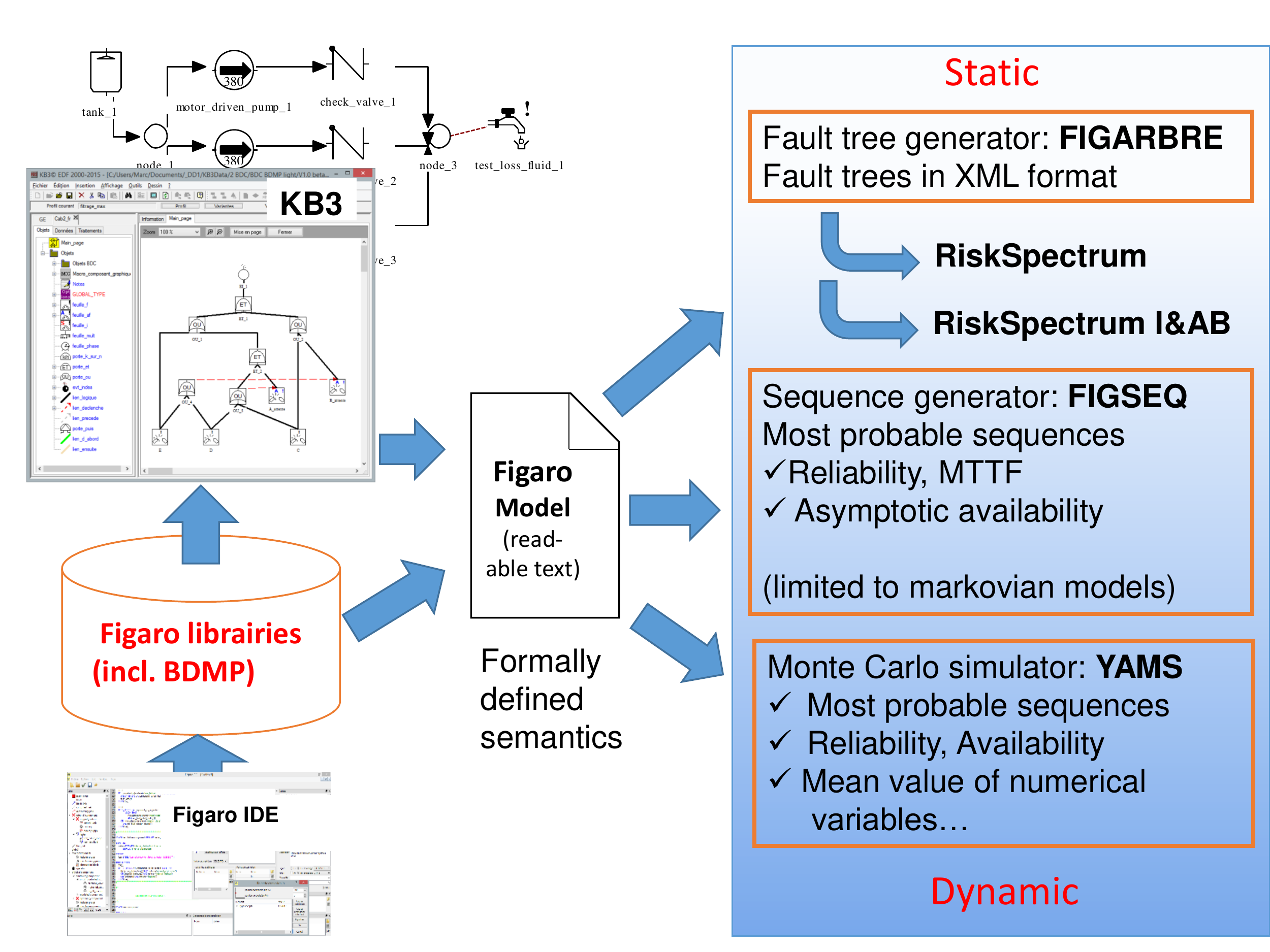}}
\caption{The KB3 workbench.}
\label{fig:kb3}
\end{wrapfigure}

The theoretical concepts of BDMP could be implemented easily in powerful software tools thanks to the \kb workbench~\footnote{\href{http://sourceforge.net/projects/visualfigaro/files/VisualFigaro/}{http://sourceforge.net/projects/visualfigaro/files/VisualFigaro/}} \cite{bouissou2005yamsfigseq} based on the Figaro modeling language \cite{bouissouSafecomp1991}.
Figaro is object-oriented: a system model is made of a set of hierarchically organized classes constituting a ``knowledge base" or library, and objects inheriting their characteristics (state variables, stochastic transitions and deterministic propagation of interactions) from the classes. Figure~\ref{fig:kb3} shows the architecture of the workbench: 

\begin{enumerate}
\item	Knowledge bases in Figaro, containing generic models for a category of systems like thermohydraulic systems or abstract objects like the leaves, gates etc.\ of a BDMP are developed by knowledge management experts using a specialized editor: FigaroIDE~\footnote{\href{http://ariste.fr/figaroide.html}{http://ariste.fr/figaroide.html}};
\item	The user loads a knowledge base in the tool \kb which becomes a dedicated GUI and then builds a system model with this GUI;
\item	The system model is then translated into a purely textual representation, in the Figaro 0 language, a sub-language of Figaro that enables formal verifications and transformations;
\item	The Figaro 0 model can either be transformed into a fault tree and sent to \riskspecPSA or another fault tree processing tool or be directly processed by one of the two dynamic model solvers \figseq or \yams. Depending on the characteristics of the Figaro 0 model, only a sub-set of these three uses may be available.
\end{enumerate}

The BDMP knowledge base not only implements all concepts of \cite{bouissou2003new}: it also includes generalized stochastic Petri nets (GSPNs). It is possible to build models essentially made of a BDMP, in interaction with one or several GSPNs. The use of such extensions gives the tool KB3-BDMP (KB3 in which the BDMP knowledge base is loaded) a very high modeling power and flexibility, but it may reduce the number of possible solvers.

\figseq explores paths in the Markov graph implicitly defined by the model whose behaviour is described in a Figaro 0 file. The model must be Markovian, which means that only two kinds of transitions are allowed: instantaneous probabilistic choices and timed transitions associated to an exponentially distributed delay. 
Most systems, even very reliable ones, have some weak points. The exploration of all sequences having a probability larger than a given threshold can reveal those weak points and save the work of exploration of sequences negligible compared to dominant sequences. Figseq always gives an upperbound for the \emph{total} probability of discarded sequences, hence the use of cutoff criteria is safe.    

But there is an additional advantage if the model to be processed is a BDMP. It is the properties of BDMP that reduce drastically the number of sequences to explore thanks to the “relevant event filtering” technique, explained in detail in \cite{bouissou2003new}. Here is an intuitive summary of the relevant event filtering: supposing that a subsystem is failed, it is both closer to reality and good for reducing the state space size to inhibit all failures in this subsystem, until it is repaired. Then all sequences explored by Figseq only contain relevant events, i.e., events that make the system state closer to a global failure. 

\yams uses event driven, non-accelerated Monte Carlo simulation. 
Besides simplicity, the main advantage of this method is that it can process any Figaro 0 model, whatever the probability distributions associated to transitions (exponential, fixed time, uniform, Weibull, Lognormal, Gamma, Beta etc.).
The main drawback of Monte Carlo simulation is its large CPU consumption and its lack of precision in the case of reliable systems.

Since 2003, KB3-BDMP has been used with \figseq and \yams for assessing the reliability, availability, and safety of many complex reconfigurable systems of various kinds: electrical, thermohydraulic, information systems, etc. The possibility to quantify them with the I\&AB method in \riskspecPSA will widen their application domain as it will lift the scalability limits of a dynamic analysis. The I\&AB method can handle full-scale PSAs of nuclear power plants~\cite{bouissouPSAM2018}. 

\subsection{RiskSpectrum PSA}
\label{subsec:riskspectrum}
\riskspecPSA is a professional tool for constructing, maintaining and solving fault tree/event tree models of system reliability. 
The tool allows for full-scale Probabilistic Safety Assessment (PSA) of entire nuclear power plants. It is licensed for use at more than 60 \% of the world’s nuclear power plants. 
The calculation engine \emph{RSAT}~\cite{RiskSpectrumTheoryManual}, based on the classical MOCUS algorithm~\cite{fussell72new}, is heavily optimised and can handle Boolean models built by combining event trees and fault trees containing tens of thousands of basic events, and resulting in hundreds of thousands of minimal cut sets, while using reasonable computational resources. Appropriate use of a cutoff is instrumental in the efficiency for real-life PSA models while keeping the effect of discarding cut sets within acceptable bounds. 

All I\&AB quantification presented in this paper are made with an alpha version of \riskspecPSA / \emph{RSAT} which includes the I\&AB quantification method as an add-on. The I\&AB method decomposes the fault tree into minimal cut sets and quantifies these minimal cut sets by the calculation method sketched in Section \ref{sec:InABmethd}, taking repairs into account. For industrial-size models which have to be solved with a cutoff, it is not sufficient to use the same approach during the generation of minimal cut sets as in the static analysis. The final cut set value is not a simple product of values from the contributing basic events. This means that for a correct application of the cutoff, a conservative estimate of the value of a partially-generated cut set needs to be calculated each time we would like to apply the cutoff. Therefore, a close integration of the I\&AB method to the whole cut set generation and quantification process is needed. 

The I\&AB quantification of minimal cut sets (MCS) lists is significantly more complex than for static PSA quantification (which uses a simple product of basic event values). On the other hand, compared to a fully dynamic analysis, the required computation time for a MCS list (re-)quantification becomes very short.
This brings another advantage of this method, namely a possibility of efficient \emph{importance} and \emph{uncertainty} analyses in a dynamic setting. These calculations re-quantify the MCS list, which is generated only once, for each alteration of the model.

\section{Use case: emergency power supply of a nuclear power plant}
\label{sec:casestudy}

\subsection{System to be Studied}
\label{subsec:casestudy}
This use case, described in full details in \cite{bouissou2017benchmark} is a discrete system, including most difficulties one can encounter when assessing the dependability of a complex system: re-configurations with cascades of instantaneous probabilistic transitions, repairs, high redundancy level, common cause failures, large differences between the lowest and highest transition rates, multi-directional interactions (because of short circuits), looped interactions, existence of deterministic delays (due to battery depletion). 

This system is a part of a French nuclear power plant.
The hypotheses on the qualitative behavior of the components and the control system are as accurate as possible, but the reliability data (failure and repair rates) are fake, for confidentiality reasons. 
However, the orders of magnitude are realistic.
\subsection{Model Characteristics}
\label{subsec:modelxtics}
We used the BDMP described in \cite{bouissou2017benchmark} as the starting point of all our reliability calculations. 
Two kinds of descriptions are available for this BDMP: the graphical representation given in the article itself and the Figaro 0 file which is the basis for calculations by \figseq or \yams. 
Both the article and the model can be downloaded from the MARS 2017 workshop proceedings~\footnote{\href{http://arxiv.org/abs/1703.06575}{http://arxiv.org/abs/1703.06575}}. Here, we will just give a few characteristics of the BDMP: it comprises 81 basic events, 14 AND gates, 5 PAND (priority AND) gates and 35 OR gates. The number of states of the CTMC specified by this BDMP is unknown, but its order of magnitude is $2^{81} \approx 2.42 E^{24}$. 

We made a few simplifying assumptions in order to keep the model relatively simple. 
We supposed that the sources of the low voltage part (which, in fact, come from the high voltage part except for the battery) are perfectly reliable: this is to avoid a loop in the structure of the BDMP.
We did not model ``negligible" short-circuit propagations. 
The idea is the following: If a short circuit is possible (directly) on a bus bar, taking into account, in addition, the propagation of a short circuit coming from one of its two or three neighbors will not have a significant impact. This propagation would require the refuse to open a circuit-breaker, an event with a low probability ($< 0.001$); hence, for a bus bar $B$ the occurrence rate of "short circuit on a neighbor $and$ propagation on $B$" is negligible compared to the rate of "short circuit on $B$".
Slightly different variants of the BDMP are necessary for different uses, because of the intrinsic limitations of methods. 
The fact of having variants is easy to handle with \kb because the tool has built-in functions for this kind of task. 

In a first variant (that we call ``Markov approx." in the comparison of results), we replaced the fixed time (one hour) of the battery depletion by an Erlang distribution: Erlang(2, 2/h).
This is necessary to keep a Markovian model and to be able to use the tool \figseq which uses analytical formulae. For the Monte Carlo simulation performed with \yams, this approximation is not necessary. More precisely, two options are possible in the model, corresponding to an optimistic and a pessimistic hypothesis.
Optimistic: as soon as the battery is no longer needed, it recovers its full capacity so that it can again provide power for 1 hour.
Pessimistic: if the battery is intermittently used, it will cease to provide power as soon as the accumulated function time reaches 1 hour. This even holds for very distant uses of the battery, for example after two different initiators.
For the system studied here, this distinction makes no difference on the results, because the influence of the battery is very low.
A long repair time (1000h) is chosen for the battery, to ensure that, after a given initiating event, the battery is not restored.

In the model to be quantified by \figseq or \yams, it is possible to take into account the fact that the house load functioning cannot be repaired until the GRID is available.
To do so, we created a single repairman who repairs the basic events relative to the GRID and the house load functioning.
Since from the initial state, the house load functioning is not active, this repairman will always be taken first to repair the GRID.
This will inhibit the repair of the house load functioning until the GRID is repaired. With I\&AB, it is not possible to take into account this kind of dependence.
So, we set the repair rate of the two basic events corresponding to the loss on-demand and in the operation of the house load functioning to 0.
This amounts to considering that the repair is impossible after a given occurrence of an initiating event that triggers the house load functioning and that as soon as the initiating event is repaired, the house load functioning is available anew. 
\subsection{Comparison of Quantification Methods}
\label{subsec:comparison}

\subsubsection{Repairable} 
\label{subsec:comparison:Repariable}
Table~\ref{tab:1} compares the performances in terms of CPU consumption and precision of three methods, applied to the same (except for the variants mentioned above) BDMP model.
All calculations were performed on the same laptop, with an Intel Core i5 processor.

\yams is supposed to give the ``most accurate" result, in the sense that it can process the model closest to the real system behaviour.
Unfortunately, it requires more than one hour to get a rough estimation (the $\pm3E^{-6}$ in the table means that the 90\% confidence interval half width is $3E^{-6}$) of the system unreliability at $10^4$ hours.
This global result is obtained with $20$ million simulations. Each simulation ending with a system failure defines a sequence; among them, so few are qualitatively identical that only $70$ sequences appeared more than once in the simulations. This is why in the table, sequences found by \yams are said to be \textbf{among} the first (i.e. most probable) ones.

The results given by \figseq are based on the model using the Markov approximation described in the previous paragraph. This approximation has a negligible impact: the best estimate by \figseq is very close to the result of \yams.
\figseq yields the most accurate qualitative results (in the model checking community, this would rather be called diagnostic feedback).
In only 20 seconds, it finds the 106 most probable sequences (or paths in the Markov chain) leading to the failure of the system. 
For I\&AB, the implementation available now in \riskspecPSA is not able to take deterministic failures into account, this is why we also used the Markov approximation. 
The calculations by \riskspecPSA are by far the fastest ones, especially with the use of a cutoff. It is then possible to obtain a precise global result and more than one thousand dominant cut sets in less than one second! The price to pay for this rapidity is a loss of precision: the results are excessively conservative. 

\begin{table}[htbp]
\caption{Comparison of various quantification methods on the benchmark}
\resizebox{\textwidth}{!}{
\centering
\begin{tabular}{|l|c|c|c|c|}
\hline
\multicolumn{1}{|c|}{\textbf{Calculation}} & \textbf{CPU} & \textbf{Cutoff Prob.} & \textbf{Unrel. at $10^4$} & \textbf{Qualitative}\\
\multicolumn{1}{|c|}{\textbf{type}} & \textbf{Time} & \textbf{at $10^4$ h}& \textbf{best estimate} &\textbf{\#cut sets/sequences}\\
\hline
\multirow{2}{10em}{\figseq (Markov approximation)} & $20$ s & $1E^{-8}$ & $3.48E^{-5}$ & 106 first seq.\\
& $3$ m $10$ s & $1E^{-10}$ & $3.84E^{-5}$ & $1266$ first seq.\\
\hline
\yams (batteries = exact $1$ h) & $82$ m &  & $3.80E^{-5} \pm 3E^{-6}$ & $70$ seq. \textbf{among} first.\\

\hline
\multirow{2}{10em}{\riskspec I\&AB \\{(Markov approximation)}} & $19$ s & $0$ & $1.35E^{-4}$  & $467474$ cut sets\\
& $1$ s &$1E^{-10}$ & $1.34E^{-4}$ &$1187$ first\\

\hline
\end{tabular}
\label{tab:1}
}
\end{table}

However, I\&AB gives the correct dominant scenarios as it can be seen in the result files associated to this paper. It also correctly accounts for changes in reliability data. For example, in the following \emph{sensitivity analysis} we change only the mean repair time of the loss of offsite power initiator (by a common cause failure on the two lines), dividing it by a factor $10$. The new results are in Table~\ref{tab:2}. 

\begin{table}[htbp]
\caption{Effects of a sensitivity analysis: division by 10 of the mean repair time of loss of offsite power}
\resizebox{\textwidth}{!}{
\centering
\begin{tabular}{|l|c|c|c|c|}
\hline
\multicolumn{1}{|c|}{\textbf{Calculation}} & \textbf{CPU} & \textbf{Cutoff Prob.} & \textbf{Unrel. at $10^4$} & \textbf{Qualitative}\\
\multicolumn{1}{|c|}{\textbf{type}} & \textbf{Time} & \textbf{at $10^4$ h}& \textbf{best estimate} &\textbf{\#cut sets/sequences}\\
\hline
\figseq (Markov approximation) & $8$ m & $1E^{-11}$ & $3.85E^{-6}$  & $2498$ first seq.\\

\hline
\yams (batteries = exactly $1$ h) & At least $10$ h & & &\\

\hline
\riskspec I\&AB {(Markov approx.)} & $7$ s & $1E^{-11}$ & $1.46E^{-5}$  & $2448$ first cut sets \\
\hline
\end{tabular}
\label{tab:2}}
\end{table}

The Monte Carlo simulation is no longer usable on a laptop: it would require hundreds of hours to obtain a few sequences in addition to the global reliability estimator. The ratio of the I\&AB result divided by the \figseq result changes from 3.9 to 3.3: I\&AB becomes less conservative. 
In fact, this example is not favorable to I\&AB, because the dominant sequences involve several failures that can happen only in a given sequence, whereas I\&AB considers they can happen in any order. On other examples, we could find that I\&AB gives a result much closer to the result of \figseq. One specific case is a model of a data center supply very similar to the high voltage part of the current benchmark~\cite{bouissouPSAM2018}. The main difference is that in case of a loss of offsite power, the diesel generators start at once: there is no equivalent of house load functioning. For that example of data center supply, the ratio of the equivalent failure rate found by I\&AB divided by the \figseq result is only 1.03.  

\subsubsection{Non-repairable}
\label{subsec:comparison:Non-Repariable}
As explained in Section~\ref{subsec:storm}, the \storm model checker cannot directly process BDMPs. But since it can process DFTs, a formalism suitable for non-repairable systems, a non-repairable version of the benchmark BDMP was used as a new test case. Thanks to the translation rules detailed in \cite{khanprdc} it was possible to translate it into a DFT and quantify it with \storm. This made it possible to compare \storm performances with those of \figseq, \yams and \riskspecPSA on that non-repairable model. 
Since the system is not repairable, there is no longer an equivalent failure rate for the whole system, and it becomes interesting to compute the unreliability at various mission times. The results are displayed in Table~\ref{tab1e:nonrep}.
\begin{table}[!htbp]
\caption{Comparison of various quantification methods on the non-repairable version of benchmark}
\resizebox{\textwidth}{!}{
\centering
\begin{tabular}{|l|c|c|c|c|c|c|}
\hline
\multicolumn{1}{|c|}{\textbf{Calculation}}& \multicolumn{1}{|c|}{\textbf{Mission}}&  \textbf{CPU} & \textbf{Cutoff} &  \textbf{Unreliability} & \textbf{Qualitative (No. of} & \textbf{Illustration of}\\
\multicolumn{1}{|c|}{\textbf{type}} &  \multicolumn{1}{|c|}{\textbf{time}} & \textbf{time} & \textbf{prob.} & \textbf{(upper bound)} &\textbf{sequences / cut sets)}& \textbf{computation effort}\\
\hline
\multirow{3}{9em}{\figseq (Markov approximation)} & $100$ h & $134$ s& $1E^{-12}$ & $3.448E^{-6}$ ($3.453E^{-6}$)  & $3674$ first sequences &$229432$ seq. explored\\
&  $1000$ h & $23$ s & $1E^{-9}$ & $7.986E^{-3}$ ($7.993E^{-3}$)  & $8567$  first sequences &$210720$ seq. explored\\
&  $10000$ h & $32$ s& $1E^{-7}$ & $3.59E^{-1}$ ($3.61E^{-1}$)  & $12921$  first sequences &$329656$ seq. explored \\
\hline
\multirow{3}{9em}{\yams (Markov approximation)} & $100$ h & $55$ m  &\multirow{3}{5em}{Not relevant for Monte Carlo} &$3.23E^{-6}$ ($\pm 3E^{-7}$)   & $49$ seq. \textbf{among} the first & $1E^{8}$ simulation stories\\
&$1000$ h &$83$ s  & &$8.05E^{-3}$ ($\pm2E^{-4}$)   & $149$ seq. \textbf{among} the first & $1E^{6}$ simulation stories\\
&$10000$ h &$19$ s  & &$3.59E^{-1}$ ($\pm3E^{-3}$)   & $852$ seq. \textbf{among} the first & $1E^{5}$ simulation stories\\
\hline
\multirow{3}{9em}{\storm sparse \\engine} & $100$ h & $10$ s  & \multirow{3}{5em}{acc. up to machine precision} & $3.495E^{-6}$ &\multirow{3}{7em}{Not supported by \storm yet} & \multirow{3}{9em}
    {Model Type  : CTMC \\States  : $16386$\\Transitions : $138065$
}\\
&$1000$ h  &$10.4$ s & & $7.925E^{-3}$ & & \\
&$10000$ h  &$17.2$ s & & $3.604E^{-1}$ & &  \\
\hline
\hline
\multirow{3}{9em}{\riskspec (BDD Calc. from min. content of seq.)}& $100$ h & $<1$ s  & \multirow{3}{5em}{No cutoff for MCS generation}& $2.812E^{-5}$  & $53137$ cut sets & $118$ BDD nodes\\
& $1000$ h & $<1$ s  & & $2.740E^{-2}$  & $53137$ cut sets & $148$ BDD nodes \\
& $10000$ h & $<1$ s  & & $4.071E^{-1}$  & $53137$ cut sets & $390$ BDD nodes \\
\hline
\end{tabular}
\label{tab1e:nonrep}}
\end{table}

Let us \emph{first} compare the calculations that \emph{take into account all dynamic features} of the non-repairable BDMP (above the double line in the table). 
The \storm results are always in the interval given by \figseq. This indicates that the translation rules from the BDMP to DFT worked well. \storm is both faster and more precise than \figseq but it does not give any qualitative result. It is interesting to have both types of calculation, because \storm could serve for an uncertainty propagation calculation based on repeated calculations with failure rates values sampled from distributions representing their uncertainties (this technique is available in \riskspecPSA). The calculation by Monte Carlo simulation with \yams is obviously the worst of all three techniques: long calculation times (when the probability is small), imprecise numerical result and imprecise qualitative results (the number of sequences indicated is for sequences found at least twice in the simulations).

The calculations by \riskspecPSA differ from the other tools in one important aspect. That is it solves a translation of the dynamic BDMP model into a static fault tree preserving the minimal content of sequences (as minimal cut sets). Especially, \emph{the order of events is not considered}. This leads to much higher unreliability values at all mission times.
As the mission time increases, the probability of basic events increases. The conditions for a good approximation by the sum of cut set probabilities or Min Cut Upper Bound are no longer preserved. \riskspecPSA offers a quantification algorithm for minimal cut set lists based on binary decision diagrams (the MCS BDD algorithm~\cite{backstromPSAM2014,backstromPSAM2016}) that overcomes this issue. It becomes more and more difficult to compare sequences in the CTMC (their probabilities are always additive because by construction they are mutually exclusive) with cut sets as the mission time increases. Probability of a sequence in the CTMC corresponds to the failure of the components appearing in the sequence (in a given order) and nothing else. A cut set probability corresponds to the failure of the components in the cut set in any order, whatever happens with other components. For the shortest mission time in the current benchmark ($100$ h) the comparison is easier because the ``and nothing else" of the sequence has a probability close to 1. Then only the question of the order of events remains. 

In the case of the benchmark, all predominant cut sets have 4 or 5 failures in function. The quantification of the cut set implicitly considers that they can happen in any order, but in fact, only one or two sequences are possible. For example, all first cut sets are variants of the following scenario: loss of connection between the nuclear power plant and the grid, failure of house load mode, failure of the two diesel generators A and B, failure of the ultimate backup by the generator called TAC. Such a cut set corresponds to only 2 sequences (the failures of diesel generators can happen in any order, but all other events can happen only in a given order). If there was no constraint at all on events order, the cut set would correspond to $5! = 120$ sequences. The difference would be much smaller if the dominant sequences contained only one or two failures. This shows the benefit of using dynamic methods if the order of failures plays an important role. 
Note, that sequences of events occurring in a given order in a CTMC are conceptually different from sequences in event trees. 

\section{Conclusion}
\label{sec:conc}
In this article we have compared the performances of various analytical methods and of Monte Carlo simulation on a benchmark that was published at the 2017 MARS workshop. The initial model is a BDMP of medium size. Methods based on Boolean approximations show the greatest scalability, but they cannot take the highly sequential behaviour of the model into account, which explains pessimistic unreliability estimates. Monte Carlo simulation shows poor performances on this case (high CPU consumption, imprecise determination of dominant scenarios), because the probabilities to estimate are very small. Analytical calculations based on sequences or on a matrix generated from the BDMP perform well, but their scalability is limited. Future work will be about translation of repairable BDMPs into formalisms acceptable by \storm, in order to improve that scalability.

\bibliographystyle{eptcs}
\bibliography{generic}
\end{document}